\documentclass[aps,prc,twocolumn,groupedaddress,showpacs]{revtex4}

\usepackage{graphicx}

\begin{document}

\title{ Isotopic  scaling  of heavy projectile residues  from
        the collisions of  25 MeV/nucleon $^{86}$Kr with  $^{124}$Sn, $^{112}$Sn
        and  $^{64}$Ni, $^{58}$Ni.
      }
\author{G. A. Souliotis}
\author{D. V. Shetty}
\author{M. Veselsky} 
\thanks{On leave of absence from the Institute of Physics of the Slovak Academy of Sciences, Bratislava, Slovakia.} 
\author{G. Chubarian}
\author{L. Trache}
\author{A. Keksis}
\author{E. Martin}
\author{S. J. Yennello}

\affiliation{Cyclotron Institute, Texas A\&M University, College Station, TX 77843}

\date{\today}

\nopagebreak

\begin{abstract}

The scaling  of the yields of heavy  projectile residues  from the reactions of 25
MeV/nucleon $^{86}$Kr  projectiles with  $^{124}$Sn,$^{112}$Sn and $^{64}$Ni, $^{58}$Ni
targets is studied.
Isotopically resolved yield distributions of projectile fragments in the range
Z=10--36 from these  reaction pairs were measured with the MARS recoil separator
in the angular range 2.7$^o$--5.4$^o$.
For these deep inelastic collisions, the velocities of the residues,
monotonically decreasing  with Z down to Z$\simeq$26--28,
are employed to characterize the excitation energy.
The ratios R$_{21}$(N,Z) of the yields of a given fragment (N,Z)  from  each  pair
of systems are found to 
exhibit isotopic scaling (isoscaling), namely, an exponential dependence on the 
fragment atomic number Z and neutron number N.
The isoscaling is found to occur in the residue Z range corresponding to the maximum observed
excitation energies.
The corresponding isoscaling  parameters are 
$\alpha$=0.43 and $\beta$=--0.50  for the Kr+Sn system  and $\alpha$=0.27 and $\beta$=--0.34
for the Kr+Ni system. 
For the Kr+Sn system, for which the experimental angular acceptance range lies
inside the grazing angle, isoscaling was found to occur for Z$\leq$26 and N$\leq$34.
For heavier  fragments from Kr+Sn, the parameters  vary monotonically, $\alpha$ decreasing
with Z and $\beta$ increasing with N. This variation  is found to be  related to
the evolution  towards isospin  equilibration and, as such, it  can serve  as a  tracer of the 
 N/Z equilibration process.
The present  heavy-residue data  extend  the observation  of isotopic scaling  from the 
intermediate mass fragment region  to the heavy-residue region.
Interestingly, such high-resolution mass  spectrometric data  can  provide  important
information on the role of isospin and isospin equilibration in peripheral and mid-peripheral
collisions, complementary to that  accessible from modern large-acceptance multidetector devices.

\end{abstract}



\maketitle

\section{Introduction}

The isotopic composition of nuclear reaction products contains  important information
on the role of the isospin on the reaction dynamics \cite{Bao1}.
Recently, increased  interest in the N/Z degree of freedom and its equilibration
\cite{SJY1,Bao0,SJY2,Rami}, 
as well as  in the isospin asymmetry dependent terms of the nuclear equation of state
\cite{Bao2,Serot}
has motivated  detailed measurements of the isotopic distributions of fragments
with Z$\geq$2 \cite{Xu,Laf,MV1,Ram}.
Isotopically resolved data  in the region Z=2--8 have revealed systematic trends,
but their N/Z properties are unavoidably  affected by the decay of the 
excited primary fragments.
It has recently been shown  \cite{Tsang1} that  isospin effects can be investigated
by comparing the yields of fragments from two similar reactions that differ only in the
isospin asymmetry. In this case, the effect of sequential decay of primary fragments can be 
bypassed to a large extent.
It has been revealed that for statistical  fragment production
mechanism(s), if  two reactions  occurring at the same temperature
have different isospin asymmetry,
the ratio R$_{21}$(N,Z) of the yields of a given fragment (N,Z) 
obtained from the two reactions 2 and 1 exhibits an exponential dependence on N and  Z of the form:

$$     R_{21}(N,Z) = C \exp(\alpha N + \beta Z) $$
   
where $\alpha$ and $\beta$ are the scaling parameters and C is an overall normalization constant.
This scaling behavior is called isotopic scaling or, isoscaling, \cite{Tsang1} and has been observed
 in a variety of reactions under the  conditions of statistical emission and  equal temperature
\cite{Tsang2,Tsang3,Botvina,Shetty}. (If the temperatures are different, a generalized 
isoscaling can be obtained with appropriate temperature corrections \cite{Tsang4}.) 
The isoscaling parameters  $\alpha$ and  $\beta$ are shown to contain information
on the early stages of  fragment formation (before sequential de-excitation).
It remains a matter of further investigation to what extent they may provide information
on the type and the details of the reaction mechanism(s).

Up to the present, the isoscaling phenomenon has been systematically investigated with isotopically 
resolved fragments not heavier than Z=8. 
It would be interesting to investigate the behavior of  heavier fragments
up to the region of  heavy residues. 
Heavy residues are known to comprise a large fraction of the
reaction  cross section at  the intermediate energy regime \cite{Fuchs}.
However, efficient collection and complete characterization of the residues in terms of their 
atomic number Z, mass number A, ionic charge state q and velocity requires the 
use of a magnetic  spectrometer. 
At this point, it should be noted that  the scaling behavior of heavy fragment yields from 
light particle induced reactions at relativistic energies on separated Sn targets 
has been recently reported \cite{Adam}. In \cite{Adam}, the fragment yields were measured with 
radiochemical techniques.
 With mass spectrometric techniques, however, apart from isotopically resolved yields,
the velocities of the fragments can be obtained with high resolution and can provide information
on the excitation energy of the primary fragments.

In the present work,  isotopic scaling  is investigated  for fragments obtained
in high-resolution  mass spectrometric studies of projectile fragments from the reactions of 
25 MeV/nucleon $^{86}$Kr with  $^{124,112}$Sn and $^{64,58}$Ni targets.
Yield ratios of fragments with Z=10--36 were obtained and isotopic scaling was investigated.
Using the velocities of the residues, the excitation energy of the 
hot primary fragments  was characterized. Finally, the values  of the 
isoscaling   parameters were  used to provide  isospin information on the hot emitting sources.
The paper is organized as follows. In Section II, a brief description of 
the experimental apparatus, the measurements and the data analysis is given.
In Section III, after examiming  the average  velocity and N/Z characteristics of the
fragments, the isotopic scaling of the fragment yields is studied in detail.
Finally, conclusions from the present study are summarized in Section IV.

\section{Experimental Methods and Data Analysis}

The present study was performed at the Cyclotron Institute of Texas A\&M
University. A  25 \hbox{MeV/nucleon} $^{86}$Kr$^{22+}$ beam  from 
the K500 superconducting cyclotron, with a typical current of $\sim$1 pnA, 
interacted with isotopically enriched targets of  $^{124}$Sn, $^{112}$Sn (2 mg/cm$^{2}$)
and $^{64}$Ni, $^{58}$Ni (4 mg/cm$^{2}$).
The reaction products were analyzed with  the MARS spectrometer \cite{MARS} offering
an angular acceptance of 9 msr and momentum acceptance of 4\%.
The primary beam struck the target at 4.0$^{o}$ relative to the optical 
axis of the spectrometer. 
The direct beam was collected inside the  target chamber on a  square Faraday cup lying 
outside of the angular acceptance of the  spectrometer.
Fragments were accepted in the polar angular range 2.7$^{o}$--5.4$^{o}$. 
This angular range lies inside the grazing angle of 6.5$^{o}$
of the Kr+Sn  reactions and mostly outside the grazing angle of 3.5$^{o}$
of the Kr+Ni reactions   at 25 MeV/nucleon \cite{Wilcke}.
It should be noted that the spectrometer angle setting was chosen to optimize the 
production of very neutron-rich fragments from the Kr+Sn systems  whose detailed study 
is reported elsewhere \cite{GS_PRL}.
An  Al foil (1 mg/cm$^2$) was used  to reset 
to equilibrium the ionic charge states of the projectile fragments.
MARS optics \cite{MARS} provides one  intermediate dispersive  image and a 
final achromatic image (focal plane). At the focal plane, 
the fragments were collected in a 5$\times$5 cm  two-element 
($\Delta $E, E)  Si detector telescope. 
The $\Delta$E detector was a position-sensitive Si strip detector 
of 110 $\mu$m  thickness,  whereas the E detector  was a   
single-element Si detector of 950 $\mu$m, respectively.
Time of flight was measured between two PPACs (parallel plate avalanche 
counters)
positioned at the dispersive image and at the focal plane, respectively, 
and separated by a distance of 13.2  m. 
The PPAC at the dispersive image was also  X--Y  position sensitive  and  
used  to record 
the position of the fragments. The horizontal position, along with NMR
measurements of the field of the MARS first dipole, 
was used to determine the magnetic rigidity, $B\rho$,  of the particles. 
The reaction products were characterized by an event-by-event measurement
of energy-loss, residual energy, time of flight, and magnetic rigidity. 
The response of the spectrometer/detector system 
to ions of known atomic number Z, mass number A, ionic charge q and 
velocity was calibrated using a low intensity $^{86}$Kr primary beam and
other  beams at 25 MeV/nucleon. 

The determination of the atomic number Z was based on the energy loss of the 
particles in the $\Delta E$ detector \cite{Hubert} and their velocity. 
The ionic charge $q$ of the particles entering MARS  was obtained from
the total energy E$_{tot}$=$\Delta$E + E, the velocity and 
the magnetic rigidity.
The measurements of Z and q had resolutions of 0.5 and 0.4 units (FWHM),
respectively.
Since the ionic charge must be an integer, we assigned integer
values of q for each event by putting windows ($\Delta q=0.4$) 
on each peak of the q spectrum. 
Using the magnetic rigidity and velocity measurement, the mass-to-charge 
A/q ratio  of each ion was obtained with a resolution of 0.3\%.
Combining the q determination with the A/q measurement, the mass A
was obtained as: $A = q_{int} \times A/q $ \,
(q$_{int}$ is the integer ionic charge) 
with a resolution  (FWHM) of 0.6 A units. 
Combination and  normalization of the data at the various magnetic
rigidity settings of the spectrometer (in the range 1.3--2.0 T\,m),
summation over all ionic  charge states
(with corrections applied for missing charge states \cite{Leon}),
and, finally,  normalization for beam current  and target thickness,
provided fragment yield distributions with respect to Z, A  and velocity.
Further details of the analysis procedure can be found in \cite{George}
and in previous work with heavier beams \cite{AuZr,AuPLF,GSfission}.
The yield distributions, summed over velocities,  were used to obtain
the yield ratios   $ R_{21}(N,Z) = Y_{2}(N,Z)/Y_{1}(N,Z) $ 
employed  in the present isotopic scaling studies.

\section{Results and Discussion}

Before embarking on the discussion of the yield ratios and scaling, we will first
examine the velocity and N/Z characteristics of the reaction products.
It is well established that reactions between massive nuclei around the Fermi energy
\cite{Fuchs} proceed via a deep inelastic transfer mechanism involving substantial 
nucleon exchange \cite{MV_SiSn,MV_NPA,HR1,HR2}.
This mechanism is responsible for the creation of highly excited primary fragments that 
deexcite to produce the observed fragments.
To obtain information about  the excitation energy of the primary fragments  from the present reactions,
we will first examine the correlation of the  measured velocity with the atomic number.
Fig. 1 presents the average
velocities of the fragments as a function of  Z. Closed symbols correspond to the reactions 
with the neutron-rich targets and open symbols to those with the neutron-poor targets. 
In this figure, we observe that for fragments  close to the projectile, the velocities
are slightly below that of the projectile, corresponding to very peripheral, low-excitation energy
events. A monotonic decrease of velocity with decreasing Z is observed,
indicative of lower impact parameters, higher momentum transfers, and thus higher excitation energies.


    \begin{figure}[h]                                        

   \includegraphics[width=0.47\textwidth, height=0.50\textheight ]{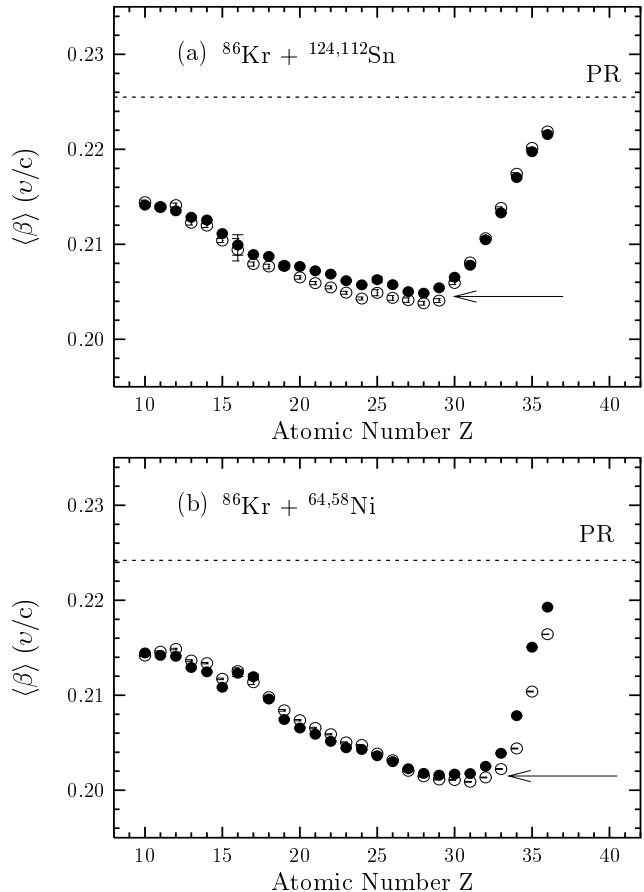}

    \caption{
           Average velocity versus atomic number Z  correlations  for projectile residues 
           from  the reactions of (a) $^{86}$Kr(25MeV/nucleon) with $^{124}$Sn and $^{112}$Sn,
           and (b)  $^{86}$Kr(25MeV/nucleon) with $^{64}$Ni and $^{58}$Ni.
	   Full circles represent the data with the neutron-rich targets $^{124}$Sn and $^{64}$Ni,
           and open circles those with  the neutron-poor  targets $^{112}$Sn and $^{58}$Ni.
           The dashed line (marked ``PR'') gives the velocity of the projectile, whereas
           the arrows indicate the minimum average residue velocities observed (see text).
           }
    \label{vel}
    \end{figure}


For the  $^{86}$Kr + $^{124,112}$Sn reactions (Fig. 1a) , the descending velocity--Z  correlation
continues  down to Z$\sim$28; for lower Z's, the velocity appears to increase with decreasing Z.
The appearance of a  minimum velocity for Z$\sim$28 can be understood by  assuming  that 
these residues originate from primary fragments with a maximum observed excitation energy. 
Fragments with Z near the projectile down to Z$\sim$28 originate from 
evaporative type of deexcitation  which does not modify, on average, the emission direction
of the residues. Thus, the residue velocities  can provide information on the excitation energy.
Residues with lower Z  arise, as we will discuss below, from 
primary  fragments undergoing cluster emission and/or multifragmentation
and the velocity of the inclusively measured fragments is not  monotonically  correlated with
the excitation energy, the mass A or the atomic number Z.
For the  $^{86}$Kr + $^{64,58}$Ni reactions, a similar behavior is observed. However, the decreasing
velocity--Z correlation is observed only down to  Z$\sim$32. It should be pointed out that fragments
from this reaction were measured mostly outside the grazing angle, so that they correspond to more
damped  collisions,  in such a way that the final residues receive a larger recoil during  the 
deexcitation stage and appear  within this angular range.
For Z$\sim$30--32, we observe a minimum velocity  and  for lower Z's an increase of the velocity
with decreasing Z. 
The average velocities from the reactions with the neutron-rich targets are, within the experimental
uncertainties, almost the same as the corresponding from the reactions with the neutron-deficient targets
for both pairs of systems.
For both reactions, the ascending  part of the velocity vs Z correlation for the lower part of the 
Z range is primarily due  to the combined effect of angle and  magnetic rigidity selection.
The forward angle range (2.7$^o$--5.4$^o$) selects  either the forward or the
backward kinematical solution in the moving frame of the quasiprojectile undergoing 
cluster-emission or multifragmentation, whereas the magnetic rigidity range (1.3--2.0 T\,m)
subsequently selects the forward solution.

Employing  the observed minimum  velocities  for the   Kr+Sn and  Kr+Ni reactions 
and, furthermore, applying  two-body  kinematics and equal division of excitation energy
(which is a reasonable assumption for nearly symmetric systems at this energy regime  
\cite{Madani,MV_SiSn}),
we can estimate an  average excitation energy per nucleon for the hot quasiprojectile 
fragments of  E$^*$/A$=$2.2$\pm$0.1  MeV
for both Kr+Sn and Kr+Ni systems. Using the Fermi gas relationship  $ E^* = \frac{A}{K} T^2 $, 
with  T  the tempretature and K the inverse level density parameter,  taken as K=13 MeV \cite{JBN1},
we can estimate the  temperature of Kr-like quasiprojectiles as T=5.3$\pm$0.2 MeV for both systems.
This  average temperature is in the range typical for reaction processes near the multifragmentation 
threshold \cite{MV1,JBN1}.
In addition, this temperature is  rather close to the limiting temperature of 
excited nuclei of  A$\sim$90 according to the recent systematics of Natowitz et al. \cite{JBN2}.
Since these  excitation energy estimates are average values obtained from average residue velocities,
we may assume  that surviving residues with even lower velocities originating  from even higher 
excitation energies are  produced in the present measurements.
However, due to the inclusive nature of the 
measurements and the velocity fluctuations due to evaporation, we cannot select  residue
source velocities  lower than those indicated by the average velocities.

In Fig. 2, we present the average N/Z values for each Z for the Kr+Sn reactions (upper panel) and
Kr+Ni reactions (lower panel). In the figures, the N/Z values from the reactions involving the
 neutron-rich targets are shown with full circles and those involving the neutron-poor targets with open
circles. The horizontal dashed line (marked ``PR'') represents the N/Z of the $^{86}$Kr projectile.
The thin solid line gives the line of $\beta$ stability (marked ``SL'') obtained from the relation:
$ Z_{\beta} = A / (1.98 + 0.0155 A^{2/3} ) $ \cite{Marmier}. 
Finally, the dotted line represents the evaporation attractor line (EAL) \cite{EAL} corresponding
to the locus of fragment yields produced by evaporation of neutrons and charged particles from
highly excited nuclei close to stability. (The statistical deexcitation code GEMINI was used to 
obtain the position of the EAL \cite{GEMINI}.)


    \begin{figure}[h]                                        

    \includegraphics[width=0.47\textwidth, height=0.50\textheight ]{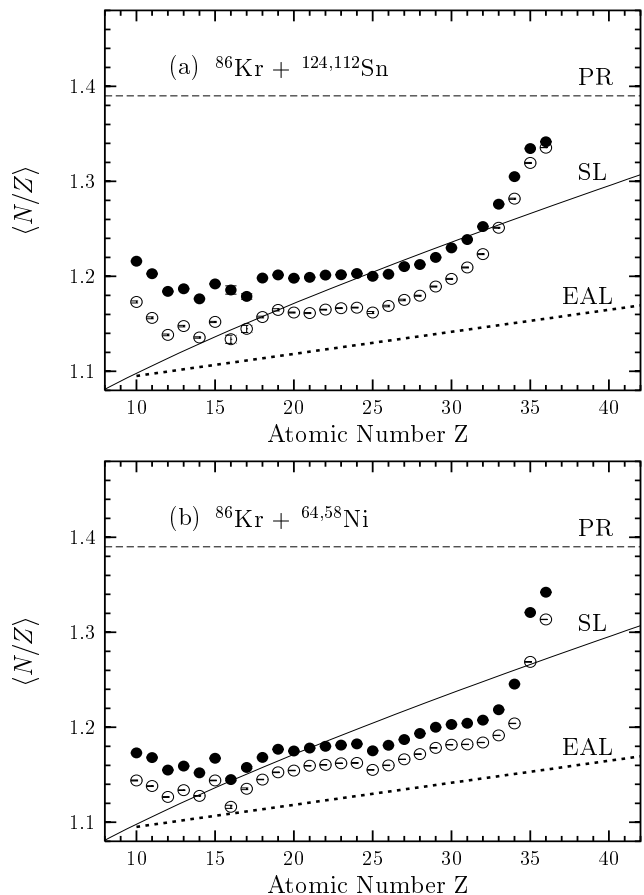}

    \caption{
           Average N/Z  versus atomic number Z  correlations  for projectile residues 
           from  the reactions of (a) $^{86}$Kr(25MeV/nucleon) with $^{124}$Sn and $^{112}$Sn,
           and (b)  $^{86}$Kr(25MeV/nucleon) with $^{64}$Ni and $^{58}$Ni.
	   Full circles represent the data with the neutron-rich targets $^{124}$Sn and $^{64}$Ni,
           and open circles those with  the neutron-poor  targets $^{112}$Sn and $^{58}$Ni.
           The dashed line (marked ``PR'') gives  the N/Z of the projectile, the solid line
           (marked ``SL'') is the line of stability and the dotted line (marked ``EAL'') is  
           the evaporation  attractor line \cite{EAL} (see text).
           }
    \label{nz}
    \end{figure}

For the Kr+Sn reactions, we observe that the average N/Z of the fragments close to the projectile
rapidly decreases with deceasing Z down to approximately Z=26--28. For lower Z,  the average 
N/Z appears to be grossly constant with a possible dip around Z=15.
The Z value below  which the average N/Z of the fragments 
is roughly constant coincides with the Z value corresponding to the minimum residue velocity,
as previously discussed in relation to Fig. 1.
As expected, the N/Z of the fragments from the reactions with the neutron-rich 
$^{124}$Sn  are larger than those from the reactions with the neutron-poor $^{112}$Sn
target.  This difference appears to be largest below Z$\sim$26, namely close to the  onset
of cluster emission and/or multifragmention.
In this region, the fragment N/Z deviates from its course towards
or beyond (to the neutron-poor side of)
the stability line  and appears to be  roughly  constant. This feature  of multifragmentation
of very neutron-rich heavy systems  may be exploited in the production of neutron-rich rare isotopes 
and it will be  the  topic of a subsequent study.  
  
For the Kr+Ni reactions, similar observations can be made in Fig. 2b. For these reactions,
the rapid decrease of N/Z is observed down to Z=32, corresponding to the Z with the minimum 
average velocity (Fig. 1b). Below Z=32, the N/Z shows a slight decrease and then,
below Z=26 it becomes roughly constant  (again with a dip at Z$\sim$15, as for Kr+Sn). 


Having examined the excitation energy and N/Z characteristics of the measured residue data,
we turn our  discussion to the scaling properties of the fragment yields.
According to the equilibrium limit of the grand-canonical ensemble, a thermally equilibrated system
undergoing statistical decay  can be characterized by a primary fragment yield with neutron number N,
and proton number Z of the form \cite{Albergo,Randrup}: 

\begin{equation}
        Y(N,Z) = F(N,Z)\exp[B(N,Z)/T]\exp(N \mu_{n}/T + Z \mu_{p}/T)
\end{equation}

where the factor $F(N,Z)$ represents contributions  due to the secondary decay from  particle stable
and  unstable states to the final ground state; $\mu_{n}$ and $\mu_{p}$ are the neutron and proton chemical 
potentials; $B(N,Z)$ is the ground state binding energy of the corresponding fragment, and $T$ is the 
temperature.
\par
A direct comparison of the observed yield distributions  with Eq. 1  is not possible  due to the 
distortions introduced by the sequential decay of the primary  fragments.
However, it has been shown  \cite{Tsang1}, that the ratio of the yields $Y_{2}(N,Z)/Y_{1}(N,Z)$  of a 
given fragment (N,Z)  from two different reacting systems,  having similar masses and excitation energies, 
but differing only  in N/Z  can be used  to obtain information about the excited primary fragments.
If the main difference between the  two systems 2 and 1   is the isospin, 
then the binding energy terms in Eq. 1 cancel out in the ratio  $Y_{2}(N,Z)/Y_{1}(N,Z)$.
Furthermore,  if one assumes that the influence of the secondary decay on the yields  is similar
for the two reactions, then a  relation of the form

\begin{equation}
     R_{21}(N,Z) = Y_{2}(N,Z)/Y_{1}(N,Z) = C \exp(\alpha N  + \beta Z)
\end {equation}       

can be obtained with  $\alpha$ = $\Delta \mu_{n}$/T and $\beta$ = $\Delta \mu_{p}$/T, with 
$\Delta \mu_{n}$ and $\Delta \mu_{p}$ being the differences in the neutron  and the 
proton chemical potentials of the fragmenting systems. C is an  overall normalization constant.
The ratio $R_{21}(N,Z)$ is insensitive to the sequential decay and thus it can provide information
on the decaying excited primary fragments.

From the present data, we construct the yield ratio $R_{21}(N,Z)$ using the
convention that index 2 refers to the more neutron-rich system and index 1 
to the less neutron-rich one.
Fig. 3 shows the yield ratios  R$_{21}$(N,Z) as a function of fragment neutron number N for 
selected isotopes
(top panel) and proton  number Z for selected isotones (bottom panel)   for the Kr+Sn reactions.
The different isotopes and isotones considered are shown by alternating filled and open symbols
for clarity. In Fig. 4, the corresponding ratios for the Kr+Ni reactions are shown.


    \begin{figure}[h]                                        

    \includegraphics[width=0.47\textwidth, height=0.50\textheight ]{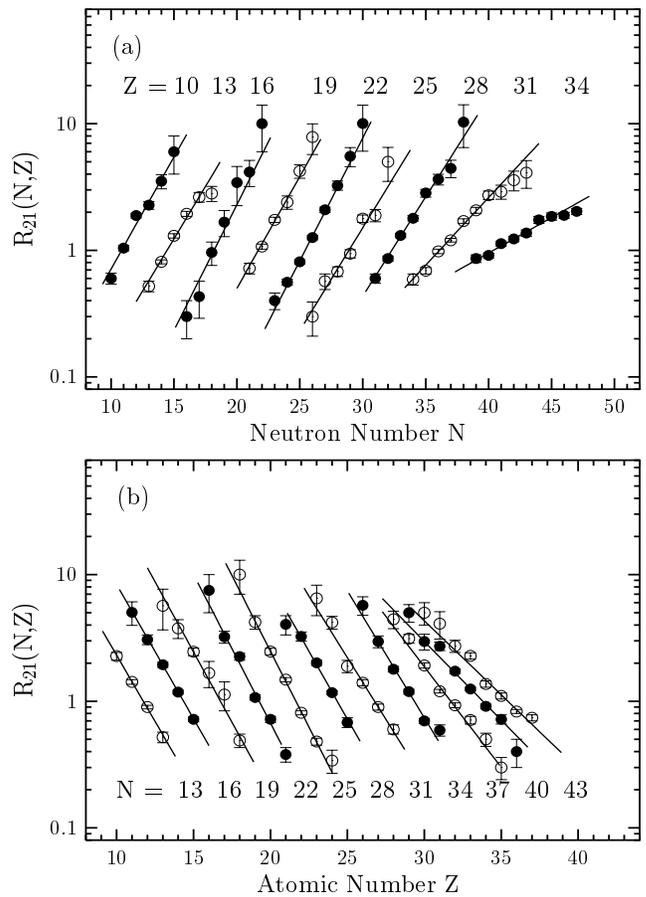}

    \caption{
           Yield ratios $ R_{21}(N,Z) = Y_{2}(N,Z)/Y_{1}(N,Z) $ of  projectile residues 
           from  the reactions of $^{86}$Kr(25MeV/nucleon) with $^{124,112}$Sn (a) with respect
           to N for the Z's indicated,  and (b) with respect to Z for the N's indicated. 
           The data are given by alternating filled and open circles, whereas the lines are
           exponential fits (see text).
           }
    \label{isoscale}
    \end{figure}


    \begin{figure}[h]                                        

    \includegraphics[width=0.47\textwidth, height=0.50\textheight ]{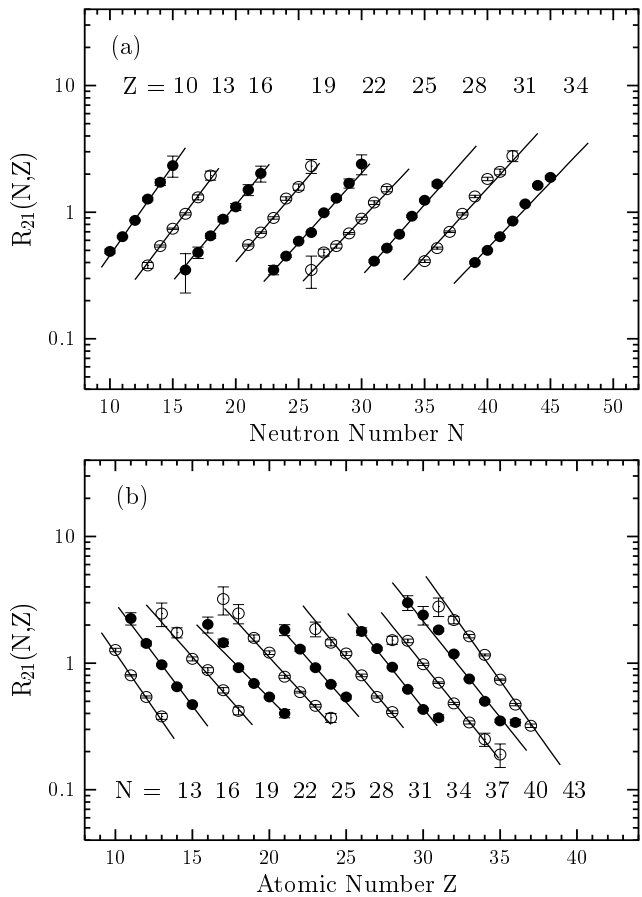}

    \caption{
           Yield ratios $ R_{21}(N,Z) = Y_{2}(N,Z)/Y_{1}(N,Z) $ of  projectile residues 
           from  the reactions of $^{86}$Kr(25MeV/nucleon) with $^{64,58}$Ni (a) with respect
           to N for the Z's indicated,  and (b) with respect to Z for the N's indicated. 
           The data are given by alternating filled and open circles, whereas the lines are
           exponential fits (see text).
           }
    \label{isoscale}
    \end{figure}

As it can be seen in the top panels  of Figs. 3 and 4,  the ratios for each element Z exhibit a remarkable
exponential behavior. For each element,  an exponential function of the form  C$exp$($\alpha$ N) was fitted
to the data and also shown in Fig. 4  for the selected elements.  In the semi-log representation, 
the lines for each element are nearly parallel up to Z$\sim$28 for Kr+Sn and up to Z$\sim$34 for Kr+Ni
(in the latter case exhausting the whole  range of observed fragments).
For heavier fragments from  the Kr+Sn systems, the fits to the data show gradual  decrease in the slopes
with increasing Z of the fragments.
An analogous behavior is observed in the lower panels  of Figs. 3 and 4, where the ratio R$_{21}$(N,Z)
is plotted as a function of  Z for various isotones. The solid lines correspond to  exponential
fits  using the expression C$exp$($\beta$Z). 
The ratios R$_{21}$(N,Z) appear  to lie along  straight lines  with nearly similar
negative slopes $\beta$, for all isotones up to N$\sim$34 for Kr+Sn
and N$\sim$44 for Kr+Ni (again, in the latter case covering the whole fragment range).
For larger N from  Kr+Sn,  the slopes $\beta$ increase   with increasing N of the fragments.
The positive  slopes  in the upper panels  of Figs. 3 and 4 indicate  that  neutron-rich 
fragments are more efficiently produced, as expected,  from the more neutron-rich systems.
Similarly, the negative slopes  in the bottom panels of these figures  indicate  that proton-rich
fragments are more efficiently produced from the more proton-rich systems.


    \begin{figure}[h]                                        

    \includegraphics[width=0.47\textwidth, height=0.50\textheight ]{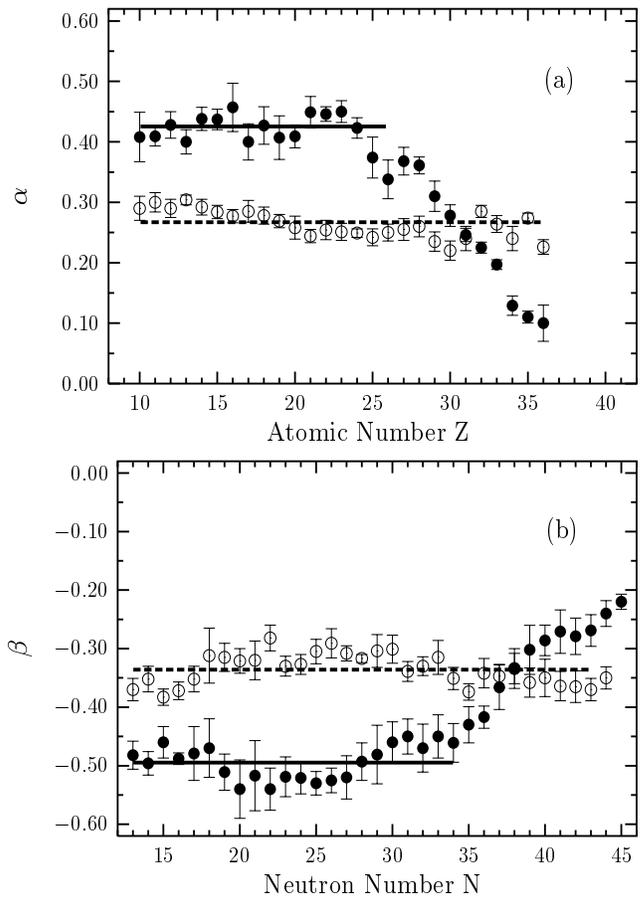}

    \caption{
           (a)  Isoscaling parameter $\alpha$ as a function of Z for 
           projectile residues from  
           the reactions $^{86}$Kr(25MeV/nucleon)+$^{124,112}$Sn
           (closed circles) and $^{86}$Kr(25MeV/nucleon)+$^{64,58}$Ni (open circles).
           The straight lines are constant value fits for each system (see text).
           (b)  Isocaling parameter $\beta$ as a function of N for 
           $^{86}$Kr(25MeV/nucleon)+$^{124,112}$Sn
           (closed circles) and $^{86}$Kr(25MeV/nucleon)+$^{64,58}$Ni (open circles).
           The straight lines are again constant value fits
           for each system.
           }
    \label{ab}
    \end{figure}

In  Fig. 5,  we present  the slope parameters $\alpha$ (upper panel) and $\beta$ (lower panel) of the
exponential  fits (obtained as described for Figs. 3 and 4)   as a function of Z and N.
For the Kr+Sn reactions, the slope parameter $\alpha$ is roughly  constant with  an average  value of
0.43 in the range Z=10--26 and then decreases  for Z$>$26.
Similarly, the parameter $\beta$ appears  to be  nearly 
constant at an average  value of --0.50  up to N$\sim$34 and  rise for  N$>$34.
Consequently, for the Kr+Sn systems  under the present experimental conditions, 
isoscaling  (as defined in \cite{Tsang1}) holds up to Z$\sim$26 and N$\sim$34.
This fragment range  corresponds to primary events with the maximum
observed excitation energy of 2.2 MeV/nucleon  and temperature of 5.3 MeV,
as estimated previously.
For the Kr+Ni reactions, the slope parameter $\alpha$ appears to be constant in the whole range
Z=10--36 at an average  value  $\alpha$=0.27.  Similarly, the parameter $\beta$  remains
constant at an average value of --0.34 in the whole N range.
In the case of the Kr+Ni systems  under the present experimental conditions (observation outside
the grazing angle),  isoscaling holds in the whole range of observed fragments,  
corresponding  to primary events having on average excitation energy of 2.2  MeV/nucleon and temperature
of 5.3 MeV. The near constancy of the isoscaling parameters $\alpha$ and $\beta$ (except for near-projectile
fragments for  the Kr+Sn system) corroborates  the  statistical nature of the fragment production  under
the excitation energy and temperature values extracted from
the minimum observed  residue velocities.
In general, the extracted isoscaling parameters $\alpha$ and $\beta$ are in agreement with the 
corresponding values reported in the literature (e.g. \cite{Tsang1,Botvina}).

In the following  discussion, we will use the isoscaling parameters herein obtained to extract information
about the hot primary fragments. As we have already mentioned, the parameters  $\alpha$ and  $\beta$ are 
related to the differences in the neutron and proton chemical potentials of the corresponding  primary 
fragmenting  systems via the relations:
$\alpha$ = $\Delta \mu_{n}$/T 
and $\beta$ = $\Delta \mu_{p}$/T.
Using the average values of the parameters $\alpha$ and  $\beta$ and the estimated temperatures, 
we can  obtain the differences in the chemical potentials,  as summarized in Table I.
As seen in the table, these values are close (apart from a difference in sign) \cite{Botvina,MV1} 
to the corresponding differences in the neutron and proton separation energies, as obtained from mass tables
\cite{Moller}. The corresponding values for the protons appear to be  remarkably close (Table I).

%
%

\begin{table}[t]                     
\caption{ Excitation energy E$^{*}$/A (MeV/nucleon), temperature T (MeV), isoscaling parameters  $\alpha$ and $\beta$,
          differences in neutron and proton  chemical potentials $\Delta \mu_n $ (MeV) and
         $\Delta \mu_p $ (MeV), differences in neutron and proton separation energies
         $\Delta S_n $  (MeV) and $\Delta S_p $  (MeV), and extracted symmetry energy coefficient C$_{sym}$  (MeV) 
         for the systems
         $^{86}$Kr (25 MeV/nucleon) + $^{124,112}$Sn, and 
         $^{86}$Kr (25 MeV/nucleon) + $^{64,58}$Ni
         (see text). Parentheses give estimated errors (one standard deviation). 
                   }
\vspace{0.5cm}
\begin{center}
\begin{tabular}{cccccccccc}
\hline
\hline
 Rxn   &  E$^{*}$/A    &    T   &  $\alpha$ & $\beta$ & $\Delta \mu_n $ & $\Delta \mu_p $ & $\Delta S_n $ & $\Delta S_p $ &  C$_{sym}$  \\   \hline


             &               &        &           &         &                 &                 &               &               &             \\

Kr+Sn        &   2.2         &  5.3   &   0.43    & --0.51  &    2.28         &   --2.65        &  --1.44       &    2.23       &  27.2       \\
             &  (0.1)        & (0.2)  &  (0.03)   &  (0.03) &   (0.18)        &    (0.19)       &               &               &  (2.2)      \\

             &               &        &           &         &                 &                 &               &               &             \\

Kr+Ni        &   2.2         &  5.3   &   0.27    & --0.34  &    1.43         &   --1.80        &  --2.75       &    1.77       &  23.1       \\
             &  (0.1)        & (0.2)   & (0.02)   &  (0.02)  &  (0.11)        &    (0.13)       &               &               &  (1.9)      \\

             &               &        &           &         &                 &                 &               &               &             \\

\hline
\hline
\end{tabular}
\end{center}
\end{table}


It has been shown 
\cite{Tsang3,Botvina} that the isoscaling parameter $\alpha$ is directly related to the coefficient  C$_{sym}$ of the 
symmetry energy term of the nuclear binding energy. The following relation has been  obtained both  in the framework of the 
grand canonical limit of the statistical multifragmentation model (SMM) \cite{Botvina} and in the expanding--emitting
source (EES) model \cite{Tsang3}:

\begin{equation}
     \alpha = 4 \, \frac {C_{sym}}{ T } ( (\frac{Z_1}{A_1})^2  -  (\frac{Z_2}{A_2})^2    )
\label{a1}
\end {equation}       

where  $Z_1$,$A_1$ and  $Z_2$,$A_2$ refer to the fragmenting quasiprojectiles from reactions 1 and 2
respectively.
Using this relation, the extracted values of $\alpha$ and T for the fragments corresponding to 
multifragmentation  events and assuming  N/Z equilibration  (see  following discussion and  Fig. 6),
we get for  C$_{sym}$  the values 
of 27.2 for Kr+Sn  and 23.1 for Kr+Ni  (Table I).  These estimates  of the symmetry-energy coefficient are in  
reasonable agreement with the standard value  C$_{sym}$ = 25 MeV \cite{Botvina}.

In principle,  Eq. \ref{a1} can serve as the basis for determining   C$_{sym}$ for expanded 
multifragmenting nuclei \cite{Tsang3}, thus probing the density dependence of the symmetry energy
term of  the nuclear equation of state. However, such experimental determination requires systematic
isotopically resolved yield measurements  with good excitation energy characterization. With  the present 
spectrometric approach, in order  to avoid 
the uncertainties in excitation energy division,  as well as in N/Z equilibration, such measurements may be
performed for symmetric pairs differing  in isospin (e.g. $^{64}$Ni+$^{64}$Ni and $^{58}$Ni+$^{58}$Ni or 
$^{124}$Sn+$^{124}$Sn  and $^{112}$Sn+$^{112}$Sn). In such cases,
the excitation energies (and temperatures) can be obtained from  the residue velocities 
and the average N/Z of the quasiprojectiles  can be taken as that  of the reacting system.


    \begin{figure}[h]                                        

    \includegraphics[width=0.47\textwidth, height=0.28\textheight ]{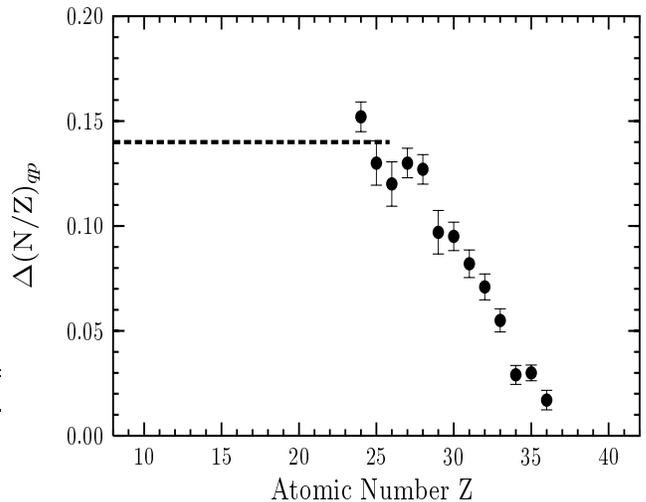}

    \caption{
               Difference in N/Z of the fragmenting quasiprojectiles (obtained  using Eq. \ref{Csym})
               as a function of Z for projectile residues from  
               the reactions  $^{86}$Kr(25MeV/nucleon)+$^{124,112}$Sn
               (closed circles).
               The dashed  line gives the N/Z difference of isospin-equilibrated 
               fragmenting quasiprojectiles.
           }
    \label{nzeq}
    \end{figure}

Finally, we will employ Eq. \ref{a1} to provide an interpretation of the variation of the isoscaling
parameter  $\alpha$ with Z
for the Kr+Sn reactions and relate this  variation to the  isospin equilibration process.
After some manipulation, from Eq. \ref{a1} we obtain:

\begin{equation}
     \alpha = 8 \, \frac{C_{sym}}{ T }  (\frac{Z}{A})_{ave}^3  \Delta( \frac{N}{Z} )_{qp}
\label{Csym}
\end {equation}       

where $(Z/A)_{ave}$ is the average Z/A of the quasiprojectiles, taken to be the average
Z/A of the composite systems $^{86}$Kr+$^{124}$Sn and $^{86}$Kr+$^{112}$Sn,  and
$ \Delta( N/Z )_{qp} $ expresses the N/Z difference of fragmenting quasiprojectiles corresponding 
to a given value of fragment Z, isoscaling parameter  $\alpha$ and temperature T.
Assuming that fragments are produced  at  normal density, 
using  C$_{sym}$ = 25 MeV 
\cite{Botvina}, the  $\alpha$ values obtained from the isoscaling fits and the temperatures
determined from excitation energies (obtained  from residue velocities), we can determine
the value  of  $ \Delta( N/Z )_{qp} $ as a function of observed fragment Z,  as shown in Fig. 6.
In this figure, the horizontal line expresses the N/Z difference of
fragmenting quasiprojectiles under the condition of  isospin equilibration.

The essentially monotonic increase  of  $ \Delta( N/Z )_{qp} $ with decreasing Z can be 
understood as follows. Fragments close to the projectile are  produced in very peripheral collisions
in which a small number of nucleons is exchanged and thus, the N/Z difference of the fragmenting 
quasiprojectiles from  $^{86}$Kr+$^{124}$Sn and $^{86}$Kr+$^{112}$Sn is small. Fragments progressively
further from the projectile originate from  collisions with larger projectile--target overlap in which 
a large number of nucleons is exchanged. In such cases,   the N/Z difference of the fragmenting 
quasiprojectiles is progressively larger,  eventually approaching, for the present energy regime, 
the N/Z difference corresponding to isospin equilibration. 
Consequently, the variation of the isoscaling parameter $\alpha$ with Z
(and,  equivalently,  $\beta$ with N) for near-projectile residues   is a result of  the variation
in the N/Z of the  fragmenting  quasiprojectiles, directly related to the course towards
isospin equilibration. As such, the parameter $\alpha$ 
along with temperature (excitation energy) determination  can be used as a sensitive probe to
study the   isospin equilibration process in collisions of isospin asymmetric massive nuclei 
at intermediate energies. 
A detailed account on this approach is reported elsewhere \cite{GS_nz}.

The present high-resolution mass spectometric data provided information on the isoscaling 
properties of heavy projectile fragments and, subsequently, information on the de-excitation
of hot primary fragments, as well as,  the degree of N/Z equilibration. Along these lines, it should be
noted that with the same spectrometric techniques,  we recently obtained information
on the details of the de-excitation mechanisms of heavy fragments and the onset of multifragmentation
in the reaction of $^{124}$Sn(25MeV/nucleon) with $^{27}$Al  \cite{MV_SnAl}.
These recent efforts in the  Fermi energy regime (following earlier work with heavier beams \cite{AuZr,AuPLF}),
along with similar studies at relativistic energies (e.g. \cite{KHS1,KHS2}  and references therein)
indicate that detailed isotopic and kinematic investigation  of  heavy residues provides  essential 
information on the reaction mechanisms and the details of the de-excitation processes that 
can complement the results of full-acceptance devices.   As a possible  next step,
combination of  high-resolution mass separators with  advanced large-acceptance multidetector arrays
may allow  novel  and exciting studies of reaction dynamics, furthermore  taking advantage of the
availability of  rare isotope beams.


\section{Summary and Conclusions}

In summary, the isotopic scaling  of heavy  projectile residues  from the interaction of 25
MeV/nucleon $^{86}$Kr  projectiles with  $^{124}$Sn,$^{112}$Sn and $^{64}$Ni, $^{58}$Ni  targets
has been  studied.
Isotopically resolved yield distributions of projectile fragments in the range
Z=10--36 from the above reaction pairs were measured with the MARS recoil separator
in the angular range 2.7$^o$--5.4$^o$.
For these deep inelastic collisions, the velocities of the residues, monotonically decreasing 
with Z down to Z$\sim$26--28, are employed to characterize the excitation energy.
The yield ratios R$_{21}$(N,Z) for each  pair  of systems are found to 
exhibit  isoscaling, namely, an exponential dependence on the fragment atomic number Z
and neutron number N.
The isoscaling is found to occur in the residue Z range corresponding to the maximum 
observed excitation energies.
The corresponding isoscaling  parameters are 
$\alpha$=0.43 and $\beta$=--0.50  for the Kr+Sn system and $\alpha$=0.27 and $\beta$=--0.34
for the Kr+Ni system. 
For the Kr+Sn system, for which the experimental angular acceptance range lies
inside the grazing angle, isoscaling was found to occur for Z$\leq$26 and N$\leq$34.
The  values of the  isoscaling parameters are used to extract the  neutron
and proton chemical potentials, as well as obtain an estimate of the symmetry energy
coefficient of highly excited  primary fragments. 
For heavy fragments from  Kr+Sn, the isoscaling behavior breaks down and the parameters are found
to vary monotonically, $\alpha$ decreasing with Z and $\beta$ increasing with N.
This variation of the isoscaling  parameters for near-projectile residues is found to be  related to
the evolution  towards isospin  equilibration and, as such can serve  as a tracer of  N/Z equilibration.
The present isotopically resolved heavy-residue
data demonstrate the  occurence of isotopic scaling  from the 
intermediate mass fragment region  to the heavy-residue region.
Such high-resolution mass  spectrometric data have the potential to  provide  important
information on the role of isospin and isospin equilibration in peripheral and mid-peripheral collisions,
complementary to that  accessible from advanced large-acceptance multidetector devices.

\section{Ackowledgements}

\par

We are thankful to M. B. Tsang  and R. T. de Souza  for insightful discussions.
We  wish to thank the Cyclotron Institute staff for the
excellent beam quality. 
This work was supported in part by the Robert A. 
Welch Foundation through grant No. A-1266, and the Department of Energy
through grant No. DE-FG03-93ER40773.  M.V. was also supported through grant
VEGA-2/1132/21 (Slovak Scientific Grant Agency).


\bibliography{isoscaling.bib}


\end{document}